# What Next?

## A Dozen Information-Technology Research Goals

Jim Gray

June 1999

Microsoft Research
Advanced Technology Division
Microsoft Corporation
One Microsoft Way
Redmond, WA  98052

# What Next?
# A Dozen Information-Technology Research Goals[1]

Jim Gray

Microsoft Research

301 Howard St.  SF, CA 94105, USA

**Abstract:** Charles Babbage's vision of computing has largely been realized.  We are on the verge of realizing Vannevar Bush's Memex.  But, we are some distance from passing the Turing Test.   These three visions and their associated problems have provided  long-range research goals for many of us.  For example, the scalability problem has motivated me for several decades. This talk defines a set of fundamental research problems that broaden the Babbage, Bush, and Turing visions.  They extend Babbage's computational goal to include highly-secure, highly-available, self-programming, self-managing, and self-replicating systems. They extend Bush's Memex vision to include a system that automatically organizes, indexes, digests, evaluates, and summarizes information (as well as a human might). Another group of problems extends Turing's vision of intelligent machines to include prosthetic vision, speech, hearing, and other senses. Each problem is simply stated and each is orthogonal from the others, though they share some common core technologies

## 1. Introduction

This talk first argues that long-range research has societal benefits, both in creating new ideas and in training people who can make even better ideas and who can turn those ideas into products.   The education component is why much of the research should be done in a university setting.    This argues for government support of long-term university research. The second part of the talk outlines sample long-term information systems  research goals.

I want to begin by thanking the ACM Awards committee for selecting me as the 1998 ACM Turing Award winner.  Thanks also to Lucent Technologies for the generous prize.

Most of all, I want to thank my mentors and  colleagues. Over the last 40 years, I have learned from many brilliant people. Everything I have done over that time has been a team effort. When I think of any project, it was Mike and Jim, or Don and Jim, or Franco and Jim, or Irv and Jim, or or Andrea and Jim, or Andreas and Jim, Dina and Jim, or Tom and Jim, or Robert and Jim, and so on to the present day.  In every case it is hard for me to point to anything that I personally did: everything has been a collaborative effort..  It has been a joy to work with these people who are among my closest friends.

More broadly, there has been a large community working on the problems of making automatic and reliable data stores and transaction processing systems. I am proud to have been part of this effort, and I am proud to be chosen to represent the entire community.  Thank you all!

---

[1] The Association of Computing Machinery selected me as the 1998 A.M. Turing Award recipient.  This is approximately the text of the talk I gave in receipt of that award.  The slides for that talk are at http://research.microsoft.com/~Gray/Talks/Turing2.ppt

## 1.1. Exponential Growth Means Constant Radical Change.

Exponential growth has been driving the information industry for the last 100 years. Moore's law predicts a doubling every 18 months. This means that in the next 18 months there will be as much new storage as all storage ever built, as much new processing as all the processors ever built. The area under the curve in the next 18 months equals the area under the curve for all human history.

In 1995, George Glider predicted that deployed bandwidth would triple every year, meaning that it doubles every 8 months. So far his prediction has been pessimistic: deployed bandwidth seems to be growing faster than that!

This doubling is only true for the underlying technology, the scientific output of our field is doubling much more slowly. The literature grows at about 15%, per year, doubling every five years.

Exponential growth cannot go on forever. E. coli (bacteria in your stomach) double every 20 minutes. Eventually something happens to limit growth. But, for the last 100 years, the information industry has managed to sustain this doubling by inventing its way around each successive barrier. Indeed, progress seems to be accelerating (see Figure 1). Some argue that this acceleration will continue, while others argue that it may stop soon – certainly if we stop innovating it will stop tomorrow.

These rapid technology doublings mean that information technology must constantly redefine itself: many things that were impossibly hard ten years ago, are now relatively easy. Tradeoffs are different now, and they will be very different in ten years.

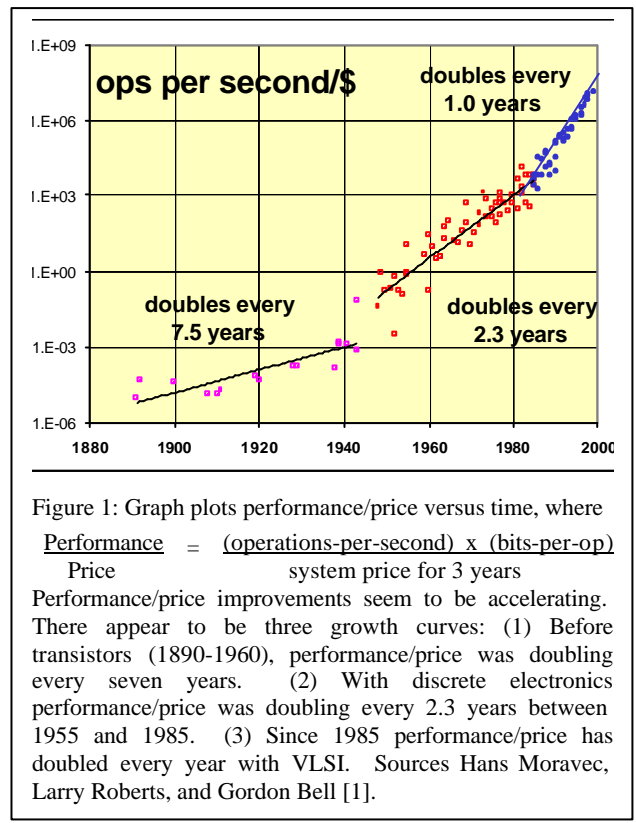

Figure 1: Graph plots performance/price versus time, where

$$\frac{\text{Performance}}{\text{Price}} = \frac{\text{(operations-per-second) x (bits-per-op)}}{\text{system price for 3 years}}$$

Performance/price improvements seem to be accelerating. There appear to be three growth curves: (1) Before transistors (1890-1960), performance/price was doubling every seven years. (2) With discrete electronics performance/price was doubling every 2.3 years between 1955 and 1985. (3) Since 1985 performance/price has doubled every year with VLSI. Sources Hans Moravec, Larry Roberts, and Gordon Bell [1].

## 1.3. Cyberspace is a *New World*

One way to think of the Information Technology revolution is to think of cyberspace as a new continent -- equivalent to discovery of the Americas 500 years ago. Cyberspace is transforming the old world with new goods and services. It is changing the way we learn, work, and play. It is already a trillion dollar per year industry that has created a trillion dollars of wealth since 1993. Economists believe that 30% of the United States economic growth comes from the IT industry. These are high-paying high-export industries that are credited with the long boom – the US economy has skipped two recessions since this boom started.

With all this money sloshing about, there is a gold rush mentality to stake out territory. There are startups staking claims, and there is great optimism. Overall, this is a very good thing.

## 1.4. This new world needs explorers, pioneers, and settlers

Some have lost sight of the fact that most of the cyberspace territory we are now exploiting was first explored by IT pioneers a few decades ago. Those prototypes are now transforming into products.

The gold rush mentality is casing many research scientists to work on near-term projects that might make them rich, rather than taking a longer term view. Where will the next generation get its prototypes if all the explorers go to startups? Where will the next generation of students come from if the faculty leave the universities for industry?

Many believe that it is time to start Lewis and Clark style expeditions into cyberspace: major university research efforts to explore far-out ideas, and to train the next generation of research scientists. Recall that when Tomas Jefferson bought the Louisiana Territories from France, he was ridiculed for his folly. At the time, Jefferson predicted that the territories would be settled by the year 2000. To accelerate this, he sent out the Lewis & Clark expedition to explore the territories. That expedition came back with maps, sociological studies, and a corps of explorers who led the migratory wave west of the Mississippi [6].

We have a similar opportunity today: we can invest in such expeditions to create the intellectual and human seed corn for the IT industry of 2020. It is the responsibility of government, industry, and private philanthropy to make this investment for our children, much as our parents made this investment for us.

## 1.5. Pioneering research pays off in the long-term

To see what I mean, I recommend you read the NRC Brooks Southerland report, *Evolving the High-Performance Computing and Communications Initiative to Support the nations Information Infrastructure* [2] or more recently: *Funding the Revolution* [3]. Figure 2 is based on a figure that appears in both reports. It shows how government-sponsored and industry-sponsored research in Time Sharing turned into a billion dollar industry after a decade. Similar things happened with research on graphics, networking, user interfaces, and many other fields. Incidentally, much of this research work fits within Pasteur's Quadrant [5], IT research generally focuses on fundamental issues, but the results have had enormous impact on and benefit to society.

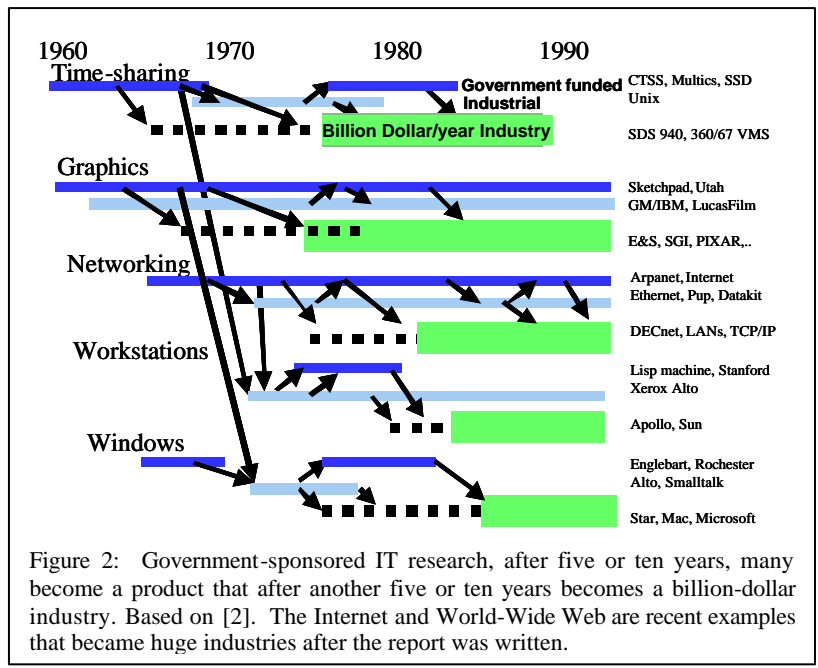

Figure 2: Government-sponsored IT research, after five or ten years, many become a product that after another five or ten years becomes a billion-dollar industry. Based on [2]. The Internet and World-Wide Web are recent examples that became huge industries after the report was written.

Closer to my own discipline, there was nearly a decade of research on relational databases before they became products. Many of these products needed much more research before they could deliver on their usability, reliability, and performance promises. Indeed, active research on each of these problems continues to this day, and new research ideas are constantly feeding into the products. In the mean time, researchers went on to explore parallel and distributed database systems that can search huge databases, and to explore data mining techniques that can quickly summarize data and find interesting patterns, trends, or anomalies in the data. These research ideas are just now creating another billion-dollar-per-year industry.

Research ideas typically need a ten year gestation period to get to products. This time lag is shortening because of the gold rush. Research ideas still need time to mature and develop before they become products.

## 1.6. Long-term research is a public good

If all these billions are being made, why should government subsidize the research for a trillion-dollar a year industry? After all, these companies are rich and growing fast, why don't they do their own research?

The answer is: "most of them do." The leading IT companies (IBM, Intel, Lucent, Hewlett Packard, Microsoft, Sun, Cisco, AOL, Amazon,...) spend between 5% and 15% of their revenues on Research and Development. About 10% of that R&D is not product development. I guess about 10% of that (1% of the total) is pure long-term research not connected to any near term product (most of the R of R&D is actually advanced development, trying to improve existing products). So, I guess the IT industry spends more than 500 million dollars on long-range research, which funds about 2,500 researchers. This is a conservative estimate, others estimate the number is two or three times as large. By this conservative measure, the scale of long-term industrial IT research is comparable to the number of tenure-track faculty in American computer science departments.

Most of the IT industry does fund long-range IT research; but, to be competitive some companies cannot. MCI-WorldCom has no R&D line item in the annual report, nor does the consulting company EDS. Dell computer has a small R&D budget. In general, service companies and systems integrators have very small R&D budgets.

One reason for this is that long-term research is a social good, not necessarily a benefit to the company. AT&T invented the transistor, UNIX, and the C and C++ languages. Xerox invented Ethernet, bitmap printing, iconic interfaces, and WYSIWYG editing. Other companies like Intel, Sun, 3Com, HP, Apple, and Microsoft got the main commercial benefits from this research. Society got much better products and services -- that is why the research is a public good.

Since long-term research is a public good, it needs to be funded as such: making all the boats rise together. That is why funding should come in part from society: industry is paying a tax by doing long-term research; but, the benefits are so great that society may want to add to that, and fund university research. Funding university research has the added benefit of training the next generations of researchers and IT workers. I do not advocate Government funding of industrial research labs or government labs without a strong teaching component.

One might argue that US Government funding of long-term research benefits everyone in the world. So why should the US fund long-term research? After all, it is a social good and the US is less than 10% of the world. If the research will help the Europeans and Asians and Africans, the UN should fund long-term research.

The argument here is either altruistic or jingoistic. The altruistic argument is that long-term research is an investment for future generations world-wide. The jingoistic argument is that the US leads the IT industry. US industry is extremely good at transforming research ideas into products – much better than any other nation.

To maintain IT leadership, the US needs people (the students from the universities), and it needs new ideas to commercialize. But, to be clear, this is a highly competitive business, cyberspace is global, and the workers are international. If the United States becomes complacent, IT leadership will move to other nations.

## 1.7. The PITAC report and its recommendations.

Most of my views on this topic grow out of a two year study by the Presidential IT Advisory Committee (PITAC) http://www.ccic.gov/ac/report/ [4]. That report recommends that the government sponsor Lewis and Clark style expeditions to the 21$^{st}$ century, it recommends that the government double university IT research funding – and that the funding agencies shift the focus to long-term research. By making larger and longer-term grants, we hope that university researchers will be able to attack larger and more ambitious problems.

It also recommends that we fix the near-term staff problem by facilitating immigration of technical experts. Congress acted on that recommendation last year: adding 115,000 extra H1 visas for technical experts. The entire quota was exhausted in 6 months: the last FY99 H1 visas were granted in early June.

# 2. Long Range IT Systems Research Goals

Having made a plea for funding long-term research. What exactly are we talking about? What are examples of long-term research goals that we have in mind? I present a dozen examples of long-term systems research projects. Other Turing lectures have presented research agendas in theoretical computer science. My list complements those others.

## 2.1. What Makes a Good Long Range Research Goal?

Before presenting my list, it is important to describe the attributes of a good goal. A good long-range goal should have five key properties:

**Understandable**: The goal should be simple to state. A sentence, or at most a paragraph should suffice to explain the goal to intelligent people. Having a clear statement helps recruit colleagues and support. It is also great to be able to tell your friends and family what you actually do.

**Challenging**: It should not be obvious how to achieve the goal. Indeed, often the goal has been around for a long time. Most of the goals I am going to describe have been explicit or implicit goals for many years. Often, there is a camp who believe the goal is impossible.

**Useful**: If the goal is achieved, the resulting system should be clearly useful to many people -- I do not mean just computer scientists, I mean people at large.

**Testable**: Solutions to the goal should have a simple test so that one can measure progress and one can tell when the goal is achieved.

**Incremental**: It is very desirable that the goal has intermediate milestones so that progress can be measured along the way. These small steps are what keep the researchers going.

## 2.2. Scalability: a sample goal

To give a specific example, much of my work was motivated by the *scalability* goal described to me by John Cocke. The goal is to devise a software and hardware architecture that scales up without limits. Now, there has to be some kind of limit: billions of dollars, or giga-watts, or just space. So, the more realistic goal is to be able to scale from one node to a million nodes all working on the same problem.

> 1. **Scalability**: Devise a software and hardware architecture that scales up by a factor for $10^6$. That is, an application's storage and processing capacity can automatically grow by a factor of a million, doing jobs faster ($10^6$x speedup) or doing $10^6$ larger jobs in the same time ($10^6$x scaleup), just by adding more resources.

Attacking the scalability problem leads to work on all aspects of large computer systems. The system grows by adding modules, each module performing a small part of the overall task. As the system grows, data and computation has to migrate to the new modules. When a module fails, the other modules must mask this failure and continue offering services. Automatic management, fault-tolerance, and load-distribution are still challenging problems.

The benefit of this vision is that it suggests problems and a plan of attack. One can start by working on automatic parallelism and load balancing. Then work on fault tolerance or automatic management. One can start by working on the 10x scaleup problem with an eye to the larger problems.

My particular research focused on building highly-parallel database systems, able to service thousands of transactions per second. We developed a simple model that describes when transactions can be run in parallel, and also showed how to automatically provide this parallelism. This work led to studies of why computers fail, and how to improve computer availability. Lately, I have been exploring very large database applications like http://terraserver.microsoft.com/ and http://www.sdss.org/.

Returning to the scaleabilty goal, how has work on scalability succeeded over the years? Progress has been astonishing, for two reasons.

 1. There has been a lot of it.

 2. Much of it has come from an unexpected direction – the Internet.

The Internet is a world-scale computer system that surprised us all. A computer system of 100 million nodes, and now merely doubling in size each year. It will probably grow to be much larger. The PITAC worries that we do not know how to scale the network and the servers. I share that apprehension, and think that much more research is needed on protocols and network engineering.

On the other hand, we do know how to build huge servers. Companies have demonstrated single systems that can process a billion transactions per day. That is comparable to all the cash transactions in the US in a day. It is comparable to all the AOL interactions in a day. It is a lot.

In addition, these systems can process a transaction for about a micro-dollar. That is, they are very cheap. It is these cheap transactions that allow free access to the Internet data servers. In essence, accesses can be paid for by advertising.

Through a combination of hardware (60% improvement per year) and software (40% improvement per year) performance and price performance have doubled every year since 1985. Several more years of this progress are in sight (see Figures 1 and 3.)

Still, we have some very dirty laundry. Computer scientists have yet to make parallel programming easy. Most of the scaleable systems like databases, file servers, and online transaction processing are embarrassingly parallel. The parallelism comes from the application. We have merely learned how to preserve it, rather

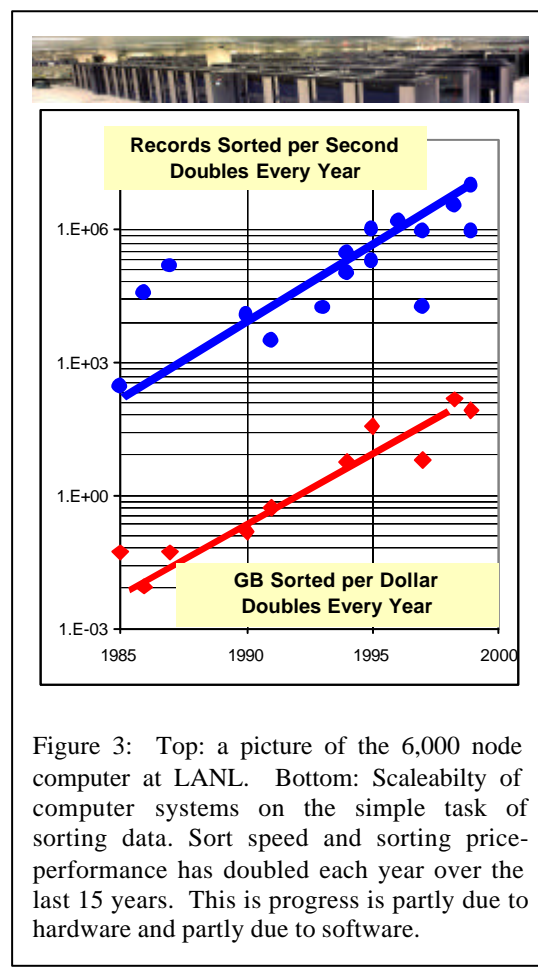

Figure 3: Top: a picture of the 6,000 node computer at LANL. Bottom: Scaleabilty of computer systems on the simple task of sorting data. Sort speed and sorting price-performance has doubled each year over the last 15 years. This is progress is partly due to hardware and partly due to software.

than creating it automatically.

When it comes to running a big monolithic job on a highly parallel computer, there has been modest progress. Parallel database systems that automatically provide parallelism and give answers more quickly, have been very successful. Parallel programming systems that ask the programmer to explicitly write parallel programs have been embraced only as a last resort. The best examples of this are the Beowulf clusters used by scientists to get very inexpensive supercomputers (http://www.beowulf.org/), and the huge ASCI machines, consisting of thousands of processors (see top of Figure 3.). Both these groups report spectacular performance, but they also report considerable pain.

Managing these huge clusters is also a serious problem. Only some of the automation postulated for scaleable systems has been achieved. Virtually all the large clusters have a custom-built management system. We will return to this issue later.

The scalability problem will become more urgent in the next decade. It appears that new computer architectures will have multiple execution streams on a single chip: so each processor chip will be an SMP (symmetric multi-processor). Others are pursuing processors imbedded in memories, disks, and network interface cards (e.g. http://iram.cs.berkeley.edu/istore/).Still another trend is the movement of processors to Micro-Electro-Mechanical Systems (MEMS). Each 10$ MEMS will have sensors, effectors, and onboard processing. Programming a collection of a million MEMS systems is a challenge [7].

So, the scaleabilty problem is still an interesting long-term goal. But, in this lecture, I would like to describe a broad spectrum of systems-research goals.

## 2. Long-term IT Systems Research Goals

In looking for the remaining eleven long-term research problems I read the previous Turing lectures, consulted many people, and ultimately settled on organizing the problems in the context of three seminal visionaries of our field. In the 1870s Charles Babbage had the vision of programmable computers that could store information and could compute much faster than people. In the 1940's Vannevar Bush articulated his vision of a machine that stored all human knowledge. In 1950, Alan Turing argued that machines would eventually be intelligent.

The problems I selected are systems problems. Previous Turing talks have done an excellent job of articulating an IT theory research agenda. Some of the problems here necessarily have an "and prove it" clause. These problems pose challenging theoretical issues. In picking the problems, I tried to avoid specific applications – trying rather to focus on the core issues of information technology that seem generic to all applications.

One area where I wish I had more to say, is the topic of ubiquitous computing. Alan Newell first articulated the vision of an intelligent universe in which every part of our environment is intelligent and networked [8]. Many of the research problems mentioned here bear on this ubiquitous computing vision, but I have been unable to crisply state a specific long-term research goal that is unique to it.

# 3. Turing's vision of machine intelligence

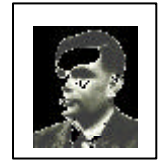

To begin, recall Alan Turing's famous "Computing Machinery and Intelligence" paper published in 1950 [9]. Turing argued that in 50 years, computers would be intelligent.

This was a *very* radical idea at that time. The debate that raged then is largely echoed today: Will computers be tools, or will they be conscious entities, having identity, volition, and free will? Turing was a pragmatist. He was just looking for intelligence, not trying to define or evaluate free will. He proposed a test, now called the *Turing Test*, that for him was an intelligence litmus test.

## *3.1 The Turing Test*

The Turing Test is based on the *Imitation Game*, played by three people. In the imitation game, a man and a woman are in one room, and a judge is in the other. The three cannot see one another, so they communicate via Email. The judge questions them for five minutes, trying to discover which of the two is the man and which is the woman. This would be very easy, except that the man lies and pretends to be a woman. The woman tries to help the judge find the truth. If the man is a really good impersonator, he might fool the judge 50% of the time. In practice, it seems the judge is right about 70% of the time.

Now, the Turning Test replaces the man with a computer pretending to be a woman. If the computer can fool the judge 30% of the time, it passes the Turing Test.

> 2. **The Turing Test:** Build a computer system that wins the imitation game at least 30% of the time.

Turing's actual text on this matter is worth re-reading. What he said was:

> *"I believe that in about fifty years' time it will be possible, to programme computers, with a storage capacity of about $10^9$, to make them play the imitation game so well that an average interrogator will not have more than 70 per cent chance of making the right identification after five minutes of questioning. The original question, "Can machines think?" I believe to be too meaningless to deserve discussion. Nevertheless I believe that at the end of the century the use of words and general educated opinion will have altered so much that one will be able to speak of machines thinking without expecting to be contradicted."*

With the benefit of hindsight, Turing's predictions read very well. His technology forecast was astonishingly accurate, if a little pessimistic. The typical computer has the requisite capacity, and is comparably powerful. Turing estimated that the human memory is between $10^{12}$ and $10^{15}$ bytes, and the high end of that estimate stands today.

On the other hand, his forecast for machine intelligence was optimistic. Few people characterize computers as intelligent. You can interview ChatterBots on the Internet (http://www.loebner.net/Prizef/loebner-prize.html) and judge for yourself. I think they are still a long way from passing the Turing Test. But, there has been enormous progress in the last 50 years, and I expect that eventually a machine will indeed pass the Turing Test. To be more specific, I think it will happen within the next 50 years because I am persuaded by the argument that we are nearing parity with the storage and computational power of simple brains.

To date, machine-intelligence has been more of a partnership with scientists: a symbiotic relationship. To give some stunning examples of progress in machine intelligence, computers helped with the proofs of several theorems (the four-color problem is the most famous example [9]), and have solved a few open problems in mathematics. It was front page news when IBM's Deep Blue beat the world chess champion. Computers help design almost everything now – they are used in conceptualization, simulation, manufacturing, testing, and evaluation.

In all these roles, computers are acting as tools and collaborators rather than intelligent machines. Vernor Vinge calls this IA (intelligence amplification) as opposed to AI [11]. These computers are not forming new concepts. They are typically executing static programs with very little adaptation or learning. In the best cases, there is a pre-established structure in which parameters automatically converge to optimal settings for this environment. This is adaptation, but it is not learning new things they way a child, or even a spider seems to.

Despite this progress, there is general pessimism about machine intelligence, and artificial intelligence (AI). We are still in AI winter. The AI community promised breakthroughs, but they did not deliver. Enough people have gotten into enough trouble on the Turing Test, that it has given rise to the expression *Turing Tar Pit* "Where everything is possible but nothing is easy." *AI complete* is short for even harder than NP complete. This is in part a pun on Turing's most famous contribution: the proof that very simple computers can compute anything computable.

Paradoxically, today it is much easier to research machine intelligence because the machines are so much faster and so much less expensive. This is the "counting argument" that Turing used. Desktop machines should be about as intelligent as a spider or a frog, and supercomputers ought to be nearing human intelligence.

The argument goes as follows. Various experiments and measures indicate that the human brain stores at most $10^{14}$ bytes (100 Terabytes). The neurons and synaptic fabric can execute about 100 tera-operations per second. This is about thirty times more powerful than the biggest computers today. So, we should start seeing intelligence in these supercomputers any day now (just kidding). Personal computers are a million times slower and 10,000 times smaller than that.

This is similar to the argument that the human genome is about a billion base pairs. 90% of it is junk, 90% of the residue is in common with chimpanzees, and 90% of that residue is in common with all people. So each individual has just a million unique base pairs (and would fit on a floppy disk).

Both these arguments appear to be true. But both indicate that we are missing something *very* fundamental. There is more going on here than we see. Clearly, there is more than a megabyte difference among babies. Clearly, the software and databases we have for our super-computers is not on a track to pass the Turing Test in the next decade. Something quite different is needed. Out-of-the-box, radical thinking is needed.

We have been handed a puzzle: genomes and brains work. But we are clueless what the solution is. Understanding the answer is a wonderful long-term research goal.

## *3.2. Three more Turing Tests: prosthetic hearing, speech, and vision.*

Implicit in the Turing Test, are two sub-challenges that in themselves are quite daunting: (1) read and understand as well as a human, and (2) think and write as well as a human. Both of these appear to be as difficult as the Turing Test itself.

Interestingly, there are three other problems that appear to be easier, but still very difficult: There has been great progress on computers hearing and identifying natural language, music, and other sounds. Speech-to-text systems are now quite useable. Certainly they have benefited from faster and cheaper computers, but the algorithms have also benefited from deeper language understanding, using dictionaries, good natural language parsers, and semantic nets. Progress in this area is steady, and the error rate is dropping about 10% per year. Right now unlimited-vocabulary, continuous speech with a trained speaker and good microphone recognizes about 95% of the words. I joke that computers understand English much better than most people (note: most people do not understand English at all.) Joking aside, many blind, hearing impaired, and disabled people use speech-to-text and text-to-speech systems for reading, listening, or typing.

Speaking as well as a person, given a prepared text, has received less attention than the speech recognition problem, but it is an important way for machines to communicate with people.

There was a major thrust in language translation in the 1950s, but the topic has fallen out of favor. Certainly simple language translation systems exist today. A system that passes the Turing Test in English, will likely have a very rich internal representation. If one teaches such a system a second language, say Mandarin, then the computer would likely have a similar internal representation for information in that language. This opens up the possibility for faithful translation between languages. There may be a more direct path to good language translation, but so far it is not obvious. Bablefish (http://babelfish.altavista.com/) is a fair example of the current state of the art. It translates context-free sentences between English and French, German, Italian, Portuguese, and Spanish. It translates the sentence "Please pass the Turing Test" to "Veuillez passer l'essai de Turing", which translates back to "Please pass the test of Turing."

The third area, is visual recognition: build a system that can identify objects and recognize dynamic object behaviors in a scene (horse-running, man-smiling, body gestures,...).

Visual rendering is an area where computers already outshine all but the best of us. Again, this is a man-machine symbiosis, but the "special effects" and characters of from Lucasfilm and Pixar are stunning. Still, the challenge remains to make it easy for kids and adults to create such illusions in real time for fun or to communicate ideas.

The Turing Test also suggests prosthetic memory, but I'll reserve that for Bush's section. So the three additional Turing Tests are:

> 3. **Speech to text:** Hear as well as a native speaker.
> 4. **Text to speech:** Speak as well as a native speaker.
> 5. **See as well as a person**: recognize objects and behavior.

As limited as our current progress is in these three areas, it is still a boon to the handicapped and in certain industrial settings. Optical character recognition is used to scan text and speech synthesizers read the text aloud. Speech recognition systems are used by deaf people to listen to telephone calls and are used by people with carpal tunnel syndrome and other disabilities to enter text and commands. Indeed, some programmers use voice input to do their programming.

For a majority of the deaf, devices that couple directly to the auditory nerve could convert sounds to nerve impulses thereby replacing the eardrum and the cochlea. Unfortunately, nobody yet understands the coding used by the body. But, it seems likely that this problem will be solved someday.

Longer term these prosthetics will help a much wider audience. They will revolutionize the interface between computers and people. When computers can see and hear, it should be much easier and less intrusive to communicate with them. They will also help us to see better, hear better, and remember better.

I hope you agree that these four tests meet the criterion I set out for a good goal, they are understandable, challenging, useful, testable, and they each have incremental steps.

## 4. Bush's Memex

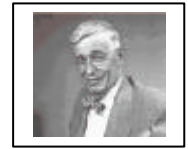

Vannevar Bush was an early information technologist: he had built analog computers at MIT. During World War II, he ran the Office of Scientific Research and Development. As the war ended he wrote a wonderful piece for the government called the *Endless Frontier [13][14]* that has defined America's science policy for the last fifty years.

In 1945 Bush published a visionary piece "As We May Think" in *The Atlantic Monthly*, http://www.theatlantic.com/unbound/flashbks/computer/bushf.htm [14]. In that article, he described *Memex*, a desk that stored "a billion books", newspapers, pamphlets, journals, and other literature, all hyper-linked together. In addition, Bush proposed a set of glasses with an integral camera that would photograph things on demand, and a Dictaphone that would record what was said. All this information was also fed into Memex.

Memex could be searched by looking up documents, or by following references from one document to another. In addition, anyone could annotate a document with links, and those annotated documents could be shared among users. Bush realized that finding information in Memex would be a challenge, so he postulated "association search", finding documents that matched some similarity criteria.

Bush proposed that the machine should recognize spoken commands, and that it type when spoken to. And, if that was not enough, he casually mentioned that "a direct electrical path to the human nervous system" might be a more efficient and effective way to ask questions and get answers.

Well, 50 years later, Memex is almost here. Most scientific literature is online The scientific literature doubles every 8 years, and most of the last 15 years are online. Much of Turing's work and Bush's articles are online. Most literature is also online, but it is protected by copyright and so not visible to the web.

The Library of Congress is online and gets more web visitors each day than regular visitors: even though just a tiny part of the library is online. Similarly, the ACM97 conference was recorded and converted to a web site. After one month, five times more people had visited the web site than the original conference. After 18 months, 100,000 people had spent a total of 50,000 hours watching the presentations on the web site. This is substantially more people and time than attendees at the actual event. The site now averages 200 visitors and 100 hours per week.

This is all wonderful, but anyone who has used the web is aware of its limitations: (1) it is hard to find things on the web, and (2) many things you want are not yet on the web. Still, the web is very impressive and comes close to Bush's vision. It is the first place I look for information. Information is increasingly migrating online to cyberspace. Most new information is created online. Today, it is about 50 times less expensive to store 100 letters (1 MB) on magnetic disk, than to store them in a file cabinet (ten cents versus 5 dollars.) Similarly, storing a photo online is about five times less expensive than printing it and storing the photo in a shoebox. Every year, the cost of cyberspace drops while the cost of real space rises.

The second reason for migrating information in cyberspace is that it can be searched by robots. Programs can scan large document collections, and find those that match some predicate. This is faster, cheaper, easier, and more reliable than people searching the documents. These searches can also be done from anywhere -- a document in England can be easily accessed by someone in Australia.

So, why isn't everything in Cyberspace? Well, the simple answer is that most information is valuable property and currently, cyberspace does not have much respect for property rights. Indeed, the cyberspace culture is that all information should be freely available to anyone anytime. Perhaps the information may come cluttered with advertising, but otherwise it should be free. As a consequence, most information on the web is indeed advertising in one form or another.

There are substantial technical issues in protecting intellectual property, but the really thorny issues revolve around the law (e.g., What protection does each party have under the law given that cyberspace is trans-national?), and around business issues (e.g., What are the economic implications of this change?). The latter two issues are retarding the move of "high value" content to the Internet, and preventing libraries from offering Internet access to their collections. Often, customers must come to the physical library to browse the electronic assets.

Several technical solutions to copy-protect intellectual property are on the table. They all allow the property owner to be paid for use of his property on a per view, or subscription, or time basis. They also allow the viewers and listeners to use the property easily and anonymously. But, until the legal and business issues are resolved, these technical solutions will be of little use.

Perhaps better schemes will be devised that protect intellectual property, but in the mean time we as scientists must work to get our scientific literature online and freely available. Much of it was paid for by taxpayers or corporations, so it should not be locked behind publisher's copyrights. To their credit, our technical society the ACM has taken a very progressive view on web publishing. Your ACM technical articles can be posted on your web site, your department's web site, and on the Computer Science Online Research Repository (CoRR). I hope other societies will follow ACM's lead on this.

## *4.1 Personal Memex*

Returning to the research challenges, the sixth problem is to build a personal Memex. A box that records everything you see, hear, or read. Of course it must come with some safeguards so that only *you* can get information out of it. But, it should on command, find the relevant event and display it to you. The key thing about this Memex is that it does not do any data analysis or summarization, it just returns what it sees and hears.

> 6. **Personal Memex:** Record everything a person sees and hears, and quickly retrieve any item on request.

Since it only records what you see and hear, personal Memex seems not to violate any copyright issues [15]. It still raises some difficult ethical issues. If you and I have a private conversation, does your Memex have the right to disclose our conversation to others? Can you sell the conversation without my permission? But, if one takes a very conservative approach: only record with permission and make everything private, then Memex seems within legal bounds. But the designers must be vigilant on these privacy issues.

Memex seems feasible today for everything but video.  A personal record of everything you ever read is about 25 GB. Recording everything you hear is a few terabytes. A personal Memex will grow at 250 megabytes (MB) per year to hold the things you read, and 100 gigabytes (GB) per year to hold the things you hear.  This is just the capacity of one modern magnetic tape or 2 modern disks.  In three years it should be one disk or tape per year.  So, if you start recording now, you should be able to stick with one or two tapes for the rest of your life.

Video Memex seems beyond our technology today, but in a few decades, it will likely be economic.  High visual quality would be hundreds times more -- 80 terabytes (TB) per year.  That is a lot of storage, eight petabytes (PB) per lifetime. It will continue to be more than most individuals can afford.  Of course, people may want very high definition and stereo images of what they see.  So, this 8 petabyte could easily rise to ten times that.  On the other hand, techniques that recognize objects might give huge image compression.  To keep the rate to a terabyte a year, the best we can offer with current compression technology is about ten TV-quality frames per second.  Each decade the quality will get at least 100x better.   Capturing, storing, organizing, and presenting this information is a fascinating long-term research goal.

## *4.2 World Memex*

What about Bush's vision of putting *all* professionally produced information into Memex?  Interestingly enough,  a book is less than a megabyte of text and all the books and other printed literature is about a petabyte in Unicode.   There are about 500,000 movies (most very short).  If you record them with DVD quality they come to about a petabyte.  If you scanned all the books and other literature in the Library of Congress the images would be a few petabytes. There are 3.5 million sound  recordings (most short) which add a few more petabytes.  So the consumer-quality digitized contents of the Library of Congress total a few petabytes. Librarians who want to preserve the images and sound want 100x more fidelity in recording and scanning the images, thus getting an exabyte.  Recording all TV and radio broadcasts (everywhere) would add 100 PB per year.

Michael Lesk did a nice analysis  of the question "How much information is there?" He concludes that there are 10 or 20 exabytes of recorded information (excluding personal and surveillance videotapes) [16]. An interesting fact is that the storage industry shipped exabyte of disk storage in 1999 and about 100 exabytes of tape storage.  Near-line (tape) and on-line (disk) storage cost between a 10 k$ and 100 k$ per terabyte.  Prices are falling faster than Moore's law – storage will likely be a hundred times cheaper in ten years.  So, we are getting close to the time when we can record most of what exists very inexpensively.  For example,  a lifetime cyberspace cemetery plot for your most recent 1 MB research report or photo of your family should cost about 25 cents.  That is 10 cents for this year, 5 cents for next year, 5 cents for the successive years, and 5 cents for insurance.

Where does this lead us?  If everything will be in cyberspace, how do we find anything?  Anyone who has used the web search engines knows both joy and frustration:  sometimes they are wonderful and find just what you want.  They do some summarization, giving title and first few sentences.  But they do very little real analysis or summarization.

So, the next challenge after a personal Memex that just returns exactly what you have seen, undigested, is a Memex that analyzes a large corpus of material and then presents it to you an a convenient way.   Raj Reddy described a system that can read a textbook and then answer the questions at the end of the text as well as a (good) college student [17]. A more demanding task is to take a corpus of text, like the Internet or the Computer Science journals, or Encyclopedia Britannica, and be able to answer summarization questions about it as well as a human expert in that field.

Once we master text, the next obvious step is to build a similar system that can digest a library of sounds (speeches, conversations, music, ...). A third challenge is a system that can absorb and summarize a collection of pictures, movies, and other imagery. The Library of congress has 115 million text and graphic items, the Smithsonian has 140 million items which are 3D (e.g. the Wright Brothers airplane). Moving those items to cyberspace is an interesting challenge. The visible humans (http://www.nlm.nih.gov/research/visible/visible_human.html), millimeter slices versions of two cadavers, give a sense of where this might go. Another exciting project is copying some of Leonardo DeVinci's work to cyberspace.

> 7. **World Memex:** Build a system that given a text corpus, can answer questions about the text and summarize the text as precisely and quickly as a human expert in that field. Do the same for music, images, art, and cinema.

The challenge in each case is to automatically parse and organize the information. Then when a someone has a question, the question can be posed in a natural interface that combines a language, gesture, graphics, and forms interface. The system should respond with answers which are appropriate to the level of the user.

This is a demanding task. It is probably AI Complete, but it an excellent goal, probably simpler and more useful than a computer that plays the imitation game as well as a human.

## *4.3 Telepresence*

One interesting aspect of being able to record everything is that other people can observe the event, either immediately, or retrospectively. I now routinely listen to lectures recorded at another research institution. Sometimes, these are "live", but usually they are on-demand. This is extraordinarily convenient -- indeed, many of us find this time-shifting to be even more valuable than space-shifting. But, it is fundamentally just television-on-demand; or if it is audio only, just radio-on-demand – turning the Internet into the world's most expensive VCR.

A much higher-quality experience is possible with the use of computers and virtual reality. By recording an event in high-fidelity from many angles, computers can reconstruct any the scene at high-fidelity from any perspective. This allows a viewer to sit anywhere in the space, or wander around the space. For a sporting event, the spectator can be on the field watching the action close up. For a business meeting, the participant can sit in the meeting and look about to read facial gestures and body language as the meeting progresses.

The challenge is to record events and then create a virtual environment on demand that allows the observer to experience the event as well as actually being there. This is called **Tele-Observer** because it is really geared to a passive observer of an event – either because it is a past event, or because there are so many observers that they must be passive (they are watching, not interacting). Television and radio give a low-quality version of this today, but they are completely passive.

The next challenge is to allow the participant to interact with the other members of the event, i.e. be **Tele-Present**. Tele-presence already exists in the form of telephones, teleconferences, and chat rooms. But, again the experience there is very much lower quality than actually being present. Indeed, people often travel long distances just to get the improved experience. The operational test for Telepresence is that a group of students taking a telepresent class score as well as students who were physically present in the classroom with the instructor. And that the instructor has the same rapport with the telepresent students, as he has with the physically present ones.

> 8. **TelePresence:** Simulate being some other place retrospectively as an observer (TeleOberserver): hear and see as well as actually being there, and as well as a participant, and simulate being some other place as a participant (TelePresent): interacting with others and with the environment as though you are actually there.

There is great interest in allowing a telepresent person to physically interact with the world via a robot. The robot is electrical and mechanical engineering, the rest of the system is information technology. That I why I have left out the robot. As Dave Huffman said: "Computer Science has the patent on the byte and the algorithm. EE has the electron and Physics has energy and matter."

## 5. Charles Babbage's Computers

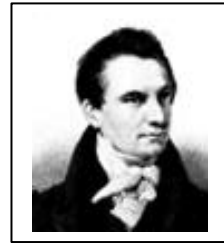

Turing's and Bush's visions are heady stuff: machine intelligence, recording everything, and telepresence. Now it is time to consider long-term research issues for traditional computers. Charles Babbage (1791-1871) had two computer designs, a difference engine, that did numeric computations well, and a fully programmable analytical engine that had punched card programs, a 3-address instruction set, and a memory to hold variables. Babbage loved to compute things and was always looking for trends and patterns. He wanted these machines to help him do his computations.

By 1955, Babbage's vision of a computer had been realized. Computers with the power he envisioned were generating tables of numbers, were doing bookkeeping, and generally doing what computers do. Certainly there is more to do on Babbage's vision. We need better computational algorithms, and better and faster machines.

But I would like to focus on another aspect of Babbage's computers. What happens when computers become free, infinitely fast, with infinite storage, and infinite bandwidth? Now, this is not likely to happen anytime soon. But, computation has gotten a 10 million times cheaper since 1950 and a trillion times cheaper since 1899 (see Figure1). Indeed, in the last decade they have gotten a thousand times cheaper. So, from the perspective of 1950, computers today are almost free, and have almost infinite speed, storage, and bandwidth.

Figure 1 charts how things have changed since Babbage's time. It measures the price-performance of these systems. Larry Roberts proposed a performance/price metric of

$$Performance/Price ? \frac{Operations\_Per\_Second ? Bits\_Per\_Operation}{System\_Price}.$$

This measures *bits-processed per dollar*. In 1969 Roberts observed that this metric was doubling about every 18 months. This was contemporaneous with Gordon Moore's observation about gates-per silicon chip doubling, but was measured at the system level. Using data from Hans' Moravac's web site (http://www.frc.ri.cmu.edu/~hpm/book98/), and some corrections from Gordon Bell, we plotted the data from Herman Hollerith forward. Between 1890 and 1945, systems were either mechanical or electro-mechanical. Their performance doubled about every seven years. In the 1950's, computing shifted to tubes and transistors and the doubling time dropped to 2.3 years. In 1985, microprocessors and VLSI came on the scene. Through a combination of lower systems prices and much faster machines, the doubling time has dropped to one year.

This acceleration in performance/price is astonishing, and it changes the rules. Similar graphs apply to the cost of storage and bandwidth.

This is real deflation. When processing, storage, and transmission cost micro-dollars, then the only real value is the data and its organization. But we computer scientists have some dirty laundry: our "best" programs typically have a bug for every thousand lines of code, and our "free" computers cost at least a thousand dollars a year in care-and-feeding known as system administration.

Computer owners pay comparatively little for them today. A few hundred dollars for a palmtop or desktop computer, a few thousand dollars for a workstation, and perhaps a few tens of thousands for a server. These folks do not want to pay a large operations staff to manage their systems. Rather, they want a self-organizing system that manages itself. For simple systems like handheld computers, the customer just wants the system to work. Always be up, always store the data, and never lose data. When the system needs repairing, it should "call home" and schedule a fix. Either the replacement system arrives in the mail, or a replacement module arrives in the mail – and no information has been lost. If it is a software or data problem, the software or data is just refreshed from the server in the sky. If you buy a new appliance, you just plug it in and it refreshes from the server in the sky (just as though the old appliance had failed).

This is the vision that most server companies are working towards in building information appliances. You can see prototypes of it by looking at WebTVs or your web browser for example.

## 5.1. Trouble-Free Systems

So, who manages the server-in-the sky? Server systems are more complex. They have some semi-custom applications, they have much heavier load, and often they provide the very services the hand-held, appliances, and desktops depend on. To some extent, the complexity has not disappeared, it has just moved.

People who own servers do not mind managing the server content, that is their business. But, they do not want to be systems management experts. So, server systems should be self managing. The human systems manager should set goals, polices, and a budget. The system should do the rest. It should distribute work among the servers. When new modules arrive, they should just add to the cluster when they are plugged in. When a server fails, its storage should have been replicated somewhere else, so the storage and computation can move to those new locations. When some hardware breaks, the system should diagnose itself and order replacement modules which arrive by express mail. Hardware and software upgrades should be automatic.

This suggests the very first of the Babbage goals: trouble-free systems.

> 9. **Trouble-Free Systems:** Build a system used by millions of people each day and yet administered and managed by a single part-time person.

The operational test for this is that it serves millions of people each day, and yet it is managed by a fraction of a person who does all the administrative tasks. Currently, such a system would need 24-hour a day coverage by a substantial staff. With special expertise required for upgrades, maintenance, system growth, database administration, backups, network management, and the like.

## 5.2. Dependable Systems

Two issues hiding in the previous requirements deserve special attention. There have been a rash of security problems recently: Melissa, Chernobyl, and now a mathematical attack on RSA that makes 512-bit keys seem dangerously small.

We cannot trust our assets to cyberspace if this trend continues. A major challenge for systems designers is to develop a system which only services authorized uses. Service cannot be denied.

Attackers cannot destroy data, nor can they force the system to deny service to authorized uses. Moreover, users cannot see data unless they are so authorized.

The added hook here is that most systems are penetrated by stealing passwords and entering as an authorized user. Any authentication based on passwords or other tokens seems too insecure. I believe we will have to go to physio-metric means like retinal scans or some other unforgeable authenticator – and that all software must be signed in an unforgeable way.

The operational test for this research goal is that a tiger team cannot penetrate the system. Unfortunately, that test does not really prove security. So this is one of those instances where the security system must rest on a *proof* that it is secure, and that all the threats are known and are guarded against.

The second attribute is that the system should always be available. We have gone from 90% availability in the 1950s to 99.99% availability today for well managed systems. Web uses experience about 99% availability due to the fragile nature of the web, its protocols, and the current emphasis on time-to-market.

Nonetheless, we have added three 9s in 45 years, or about 15 years per order-of-magnitude improvement in availability. We should aim for five more 9s: an expectation of one second outage in a century. This is an extreme goal, but it seems achievable if hardware is very cheap and bandwidth is very high. One can replicate the services in many places, use transactions to manage the data consistency, use design diversity to avoid common mode failures, and quickly repair nodes when they fail. Again, this is not something you will be able to test: so achieving this goal will require careful analysis and proof.

> 10. **Secure System:** Assure that the system of problem 9 only services authorized users, service cannot be denied by unauthorized users, and information cannot be stolen (and prove it.)
>
> 11. **AlwaysUp:** Assure that the system is unavailable for less than one second per hundred years -- 8 9's of availability (and prove it.)

### 5.3. Automatic Programming.

This brings us to the final problem: Software is expensive to write. It is the only thing in cyberspace that is getting more expensive, and less reliable. Individual pieces of software are not really less reliable, it is just that the typical program has one bug per thousand lines after it has been tested and retested. The typical software product grows fast, and so adds bugs as it grows.

You might ask how programs could be so expensive? It is simple: designing, creating, and documenting a program costs about 20$ per line. It costs about 150% of that to test the code. Then once the code is shipped, it costs that much again to support and maintain the code over its lifetime.

This is grim news. As computers become cheaper, there will be more and more programs and this burden will get worse and worse.

The solution so far is to write fewer lines of code by moving to high-level non-procedural languages. There have been some big successes. Code reuse from SAP, PeopleSoft, and others are an enormous savings to large companies building semi-custom applications. The companies still write a lot of code, but only a small fraction of what they would have written otherwise.

The user-written code for many database applications and many web applications is tiny. The tools in these areas are very impressive. Often they are based on a scripting language like JavaScript and a set of pre-built objects. Again an example of software reuse. End users are able to create impressive websites and applications using these tools.

If your problem fits one of these pre-built paradigms, then you are in luck. If not, you are back to programming in C++ or Java and producing a 5 to 50 lines of code a day at a cost of 100$ per line of code.

So, what is the solution? How can we get past this logjam? Automatic programming has been the Holy Grail of programming languages and systems for the last 45 years. Sad to report, there has been relatively little progress -- perhaps a factor of 10, but certainly not a factor of 1,000 improvement in productivity unless your problem fits one of the application-generator paradigms mentioned earlier.

Perhaps the methodical software-engineering approaches will finally yield fruit, but I am pessimistic. I believe that an entirely new approach is needed. Perhaps it is too soon, because this is a Turing Tar Pit. I believe that we have to (1) have a high level specification language that is a thousand times easier and more powerful that the current languages, (2) computers should be able to compile the language, and (3) the language should be powerful enough so that all applications can be described.

We have systems today that do any two of these three things, but none that do all three. In essence this is the imitation gave for a programming staff. The customer comes to the programming staff and describes the application. The staff returns with a proposed design. There is discussion, a prototype is built and there is more discussion. Eventually, the desired application is built.

> 12. **Automatic Programmer:** Devise a specification language or user interface that:
>     (a) makes it easy for people to express designs (1,000x easier),
>     (b) computers can compile, and
>     (c) can describe all applications (is complete).
>
> The system should reason about application, asking questions about exception cases and incomplete specification. But it should not be onerous to use.

The operational test is replace the programming staff with a computer, and produce a result that is better and requires no more time than dealing with a typical human staff. Yes, it will be a while until we have such a system, but if Alan Turing was right about machine intelligence, it's just be a matter of time.

# 6. Summary

These are a dozen very interesting research problems. Each is a long-term research problem. Now you can see why I want the government to invest in long-term research. I suspect that in 50 years future generations of computer scientists will have made substantial progress on each of these problems. Paradoxically, many (5) of the dozen problems appear to require machine intelligence as envisioned by Alan Turing.

The problems fall in the thee broad categories: Turing's intelligent machines improving the human-computer interface, Bush's Memex recording, analyzing, and summarizing everything that happens, and Babbage's computers which will finally be civilized so that they program themselves, never fail, and are safe.

No matter how it turns out, I am sure it will be very exciting. As I said at the beginning, progress appears to be accelerating: the base-technology progress in the next 18 months will equal all previous progress, if Moore's law holds. And there are lots more doublings after that.

# A Dozen Long-Term Systems Research Problems.

1. **Scalability**: Devise a software and hardware architecture that scales up by a factor for $10^6$. That is, an application's storage and processing capacity can automatically grow by a factor of a million, doing jobs faster ($10^6$x speedup) or doing $10^6$ larger jobs in the same time ($10^6$x scaleup), just by adding more resources.

2. **The Turing Test:** Build a computer system that wins the imitation game at least 30% of the time.

3. **Speech to text:** Hear as well as a native speaker.

4. **Text to speech:** Speak as well as a native speaker.

5. **See as well as a person**: recognize objects and motion.

6. **Personal Memex:** Record everything a person sees and hears, and quickly retrieve any item on request.

7. **World Memex:** Build a system that given a text corpus, can answer questions about the text and summarize the text as precisely and quickly as a human expert in that field. Do the same for music, images, art, and cinema.

8. **TelePresence:** Simulate being some other place retrospectively as an observer (TeleOberserver): hear and see as well as actually being there, and as well as a participant, and simulate being some other place as a participant (TelePresent): interacting with others and with the environment as though you are actually there.

9. **Trouble-Free Systems:** Build a system used by millions of people each day and yet administered and managed by a single part-time person.

10. **Secure System:** Assure that the system of problem 9 only services authorized users, service cannot be denied by unauthorized users, and information cannot be stolen (and prove it.)

11. **AlwaysUp:** Assure that the system is unavailable for less than one second per hundred years -- 8 9's of availability (and prove it).

12. **Automatic Programmer:** Devise a specification language or user interface that:
    (a) makes it easy for people to express designs (1,000x easier),
    (b) computers can compile, and
    (c) can describe all applications (is complete).
    The system should reason about application, asking questions about exception cases and incomplete specification. But it should not be onerous to use.